\documentclass[prd,nofootinbib,twocolumn,showpacs]{revtex4}
\usepackage{graphics}
\usepackage{bm}

\begin{document}

\title{
Bound states in bottomless potentials
}


\author{Tanmay Vachaspati}
\affiliation{Department of Physics, Case Western Reserve University,
10900 Euclid Avenue, Cleveland, OH 44106-7079, USA.}

\begin{abstract}
We consider classical and quantum dynamics on potentials that
are asymptotically unbounded from below. By explicit construction 
we find that quantum bound states can exist in certain bottomless 
potentials. The classical dynamics in these potentials is novel. 
Only a set of zero measure of classical trajectories can escape 
to infinity. All other trajectories get trapped as they get further
out into the asymptotic region. 
\end{abstract}


\

\maketitle


Certain potentials that do not have any local minima and are unbounded 
from below in the asymptotic region can still exhibit stable classical 
orbits. Specifically, consider a particle in two spatial dimensions with 
position ${\bf x}(t) = (x(t),y(t))$ and Lagrangian
\begin{equation}
L = {1\over 2} m {\bf v}^2 - V({\bf x})
\label{particleL}
\end{equation}
where
\begin{equation}
V({\bf x}) = 
  {{k_x}\over 2} x^2 - {{k_y}\over 2} y^2 + {c\over 2} x^2 y^2
\label{particleV}
\end{equation}
with $k_x$, $k_y$ and $c$ being positive parameters. The only extremum 
of the potential is at ${\bf x}=0$ and this is a saddle point. The positive
eigenvalue at the saddle point is along the $x$ direction and the
negative eigenvalue is along the $y$ direction. The potential goes
to $-\infty$ for $x^2 < k_y/c$ and $y^2 \rightarrow \infty$. Hence
the potential is ``bottomless''. Yet it is not hard to show that
there exist classical orbits that are bounded and stable. These
solutions are of the form:
\begin{equation}
x(t) = A \cos (\omega t) \ , \ \ \  y(t) =0
\label{classicalsoln}
\end{equation}
where $\omega^2 = k_x /m$ and $A$ lies in
specific bands determined by solutions of the Mathieu equation
\cite{SalVac02}.

The existence of classically bounded and stable orbits on a 
bottomless potential motivates us to consider the possibility of 
quantum bound states on a bottomless potential. Several potentials 
that are unbounded from below have already been considered in the 
quantum mechanics literature. These include the famous example of 
the Coulomb potential. Note, however, that the Coulomb potential 
is bottomless because there is a singular point where the potential 
goes to $-\infty$. In contrast, the potentials we will consider 
({\it eg.} in eq. (\ref{particleV})) go to $-\infty$ in the asymptotic 
region and are otherwise non-singular. Furthermore, we know that the 
$l=0$ wavefunctions of the hydrogen atom have discontinuous first 
derivative at the origin, while the wavefunctions with non-zero angular 
momentum vanish at the origin due to the centrifugal barrier. In our 
case, the wavefunction will be analytical throughout and its existence
cannot be attributed to a centrifugal barrier. 

The idea of the construction is quite simple. We write the two
dimensional Schrodinger equation in the form:
\begin{equation}
V({\bf x}) = E - 
 {{\hbar^2}\over {2m}}[ {\bf \nabla}^2 F - ({\bf \nabla} F )^2 ]
\label{Vxy=}
\end{equation}
in standard notation and with $F({\bf x})$ defined via
\begin{equation}
\psi = e^{- F}
\label{Fdefn}
\end{equation}
where $\psi ({\bf x})$ is the wavefunction.

We would like a solution to eq. (\ref{Vxy=}) for a bottomless
potential and a discretely normalizable wavefunction. To accomplish 
this, we choose
\begin{equation}
F = c+ \alpha x^2 + \beta y^2 + \gamma x^2 y^2
\label{Fchoice}
\end{equation}
where $\alpha$, $\beta$, and $\gamma$ are real positive 
numbers to ensure that $\psi$ is normalizable, and ${\rm exp}(-c)$
is the normalization constant. Inserting this choice of $F$ in 
eq. (\ref{Vxy=}) gives:
\begin{eqnarray}
V({\bf x})&=& \epsilon + 
      {{\hbar^2}\over m} \biggl [ (2\alpha^2-\gamma)x^2 + 
                   (2\beta^2 -\gamma) y^2  \nonumber \\
          &+& 4 (\alpha +\beta )\gamma x^2 y^2
           + 2\gamma^2 (x^2+y^2) x^2 y^2 \biggr ]
\label{Vsolution}
\end{eqnarray}
where 
\begin{equation}
\epsilon \equiv E - {{{\hbar}^2}\over m} (\alpha + \beta ) \ .
\label{Evalue}
\end{equation}
If we fix the zero of the potential to be at the origin,
this gives the energy eigenvalue of the state. Now if we 
choose $\gamma > 2 \alpha^2$ 
and/or $\gamma > 2 \beta^2$, the potential is bottomless along 
the $x$ axis and/or $y$ axis respectively. This shows an explicit
example of a bottomless potential in which there is at least
one bound state. There are many other possibilities 
that can easily be constructed in a similar way. 

There are some features of the solution that are worth pointing
out. It is easily seen that the state is not an angular momentum
($L_z = -i{\hbar}(x\partial_y - y \partial_x)$) eigenstate. 
However the expectation value of the angular momentum
operator vanishes. A peculiar feature of the potential is that it 
depends on the mass of the particle since the mass $m$ cannot be 
absorbed by rescaling the parameters $\alpha$, $\beta$ and $\gamma$. 
An interesting feature of the wavefunction is that
it has the same sign everywhere {\it i.e.} it has no nodes. 
Therefore it must be the ground state wavefunction.

One intuitive way to understand the existence of a bound
state is to examine the potential along one of the ``escape''
directions. For example, consider the potential in eq.
(\ref{Vsolution}) along the $y$ axis. In this direction ($x=0$),
the potential is falling off and getting deeper in proportion
to $- y^2$. However, the width along the $x$ direction is
also decreasing in proportion to $1/\sqrt{y^4} \propto y^{-2}$.
Hence, if we consider the $x$ dependence of the wavefunction
for large values of $y$, it corresponds to a harmonic oscillator
with angular frequency proportional to $y^2$. Therefore the
energy ``cost'' due to the squeezing in the $x$ direction
grows as $y^2$ and can be larger than the energy ``gain'' due
to rolling in the $y$ direction. This argument is basically
saying that if there is a hole in a two dimensional potential, 
quantum particles may not be able to escape through the hole if
it is sufficiently narrow. 

The quantum behaviour is in contrast to the classical
particle which can always escape by rolling along the 
$y$ axis. However, if a particle starts rolling in the $y$
direction but with $x\ne 0$, it will oscillate in the $x$
direction as it is rolling in the $y$ direction. For large
values of $y$ and small values of $x$, the equations of motion 
are:
\begin{equation}
{\ddot x} \simeq - \nu^2 y^4 x \ , \ \ 
{\ddot y} \simeq - 2 \nu^2 (x^2y^2 - \delta^2 ) y
\label{xddotyddot}
\end{equation}
where
\begin{equation}
\nu \equiv {{2\hbar}\over m}\ , \ \ 
\delta \equiv {\sqrt{\gamma - 2\beta^2 } \over {2\gamma}} \ .
\label{nudelta}
\end{equation}
Hence, the sign of the force in the $y$ direction tends to drive
the particle toward the asymptotic region only while the particle
lies in the region $x y < \delta$.
While the particle lies outside the hyperbola ({\it i.e.} when
$xy > \delta$), it experiences a restoring force which tends to
bring the particle closer to the origin. If the particle is
initially inside the hyperbola, it rolls down to larger and larger 
values of $y$, and eventually the particle orbit will increasingly 
lie outside the hyperbola. Then the particle will perform oscillations
in both the $x$ and $y$ directions at some large value of $y$. This
argument can be formalized by writing:
\begin{equation}
y(t) = Y + f(t)
\label{ypert}
\end{equation}
where $Y>0$ is a large constant value. We assume that $f(t) \ll Y$ and
that $x(t)$ remains small. Then we can do a linearized analysis in
$f(t)$ and obtain:
\begin{equation}
{\ddot x} \simeq - \nu^2 Y^4 x \ , \ \ 
{\ddot f} \simeq -2 \nu^2 (x^2 Y^2 -\delta^2 ) Y 
\label{xeqfeq}
\end{equation}
Then $x(t) = A \cos (\nu Y^2 t)$ and the $f(t)$ equation can
also be easily integrated. Periodic solutions for
$f(t)$ will be obtained when the average of the right-hand side
of the $f$ equation vanishes. This happens when $A^2 Y^2 =2\delta^2$
and the solution is:
\begin{equation}
f(t) = {{\delta^2} \over {2Y^3}} (\cos(2\nu Y t) - 1) 
\label{periodicsolution}
\end{equation}
where we have also imposed the initial conditions $f(0)=0$,
${\dot f}(0)=0$.  Hence periodic solutions do appear to leading order 
in the linearized approximation. A particle that starts rolling
but is off-axis will eventually get caught in a periodic or
quasi-periodic orbit and will not escape. 

If the above arguments are correct, it implies two interesting
corollaries. The first is that the potential in eq. (\ref{Vsolution})
does not have any continuum states since the arguments apply
to all states and no state would be able to escape to infinity. In 
principle, one could modify the potential by cutting it off at some 
large value -- that is, by setting $V({\bf x})= V_{max}$ in regions 
where the original potential (eq. (\ref{Vsolution})) exceeds 
$V_{max}$. The new potential would be bounded from above and
continuum states would then exist. The second corollary of the
above argument is that the potential in eq. (\ref{particleV}) does 
not have any bound state solutions. This is because the energy
gain along the escape path is still proportional to $y^2$ but
the energy cost due to the squeezing only grows like $\sqrt{y^2}=y$.

If we include dissipative forces, then all classical particles 
will escape to infinity since this is the lowest point on the
potential. Dissipative effects in the quantum problem can, however, 
only bring the particle into its ground state. This is just as in 
the Coulomb case where classically the atom can collapse due to 
emission of electromagnetic radiation but quantum mechanically it 
can only settle into its ground state.

Generally speaking, bottomless potentials in the field theory
context are thought to be sick. Our construction here raises the 
possibility that some field theories with bottomless potentials 
may nonetheless have reasonable interpretations.

\begin{acknowledgments} 
I am grateful to Craig Copi and Harsh Mathur for discussions.
This work was supported by DOE grant number DEFG0295ER40898 at CWRU.
\end{acknowledgments}

\end{document}